\input phyzzx\def\e{\adveq\eqno{\rm (\chapterlabel\the\equanumber)}}
\def\mysec#1{\equanumber=0\chapter{#1}}
\def\adveq{\global\advance\equanumber by 1}
\def\myeq{{\rm \chapterlabel.\the\equanumber}}
\def\rarrow{\rightarrow}

\def\semidirect{\mathrel{\raise0.04cm\hbox{${\scriptscriptstyle |\!}$
\hskip-0.175cm}\times}}

\def\mod{\mathop{\rm mod}\nolimits}

\def\ref#1{$^{[#1]}$}

\def\r#1{$[\rm#1]$}

\def\Tr{\mathop{\rm Tr}\limits}
\def\e{\adveq\eqno{\rm (\chapterlabel.\the\equanumber)}}
\def\mysec#1{\equanumber=0\chapter{#1}}
\def\adveq{\global\advance\equanumber by 1}
\def\myeq{{\rm \chapterlabel.\the\equanumber}}
\def\rarrow{\rightarrow}

\def\semidirect{\mathrel{\raise0.04cm\hbox{${\scriptscriptstyle |\!}$
\hskip-0.175cm}\times}}

\def\mod{\mathop{\rm mod}\nolimits}

\def\ref#1{$^{[#1]}$}

\def\r#1{$[\rm#1]$}

\def\Tr{\mathop{\rm Tr}\limits}

\overfullrule=0pt
\date{May, 2014}
\date{May, 2014}
\titlepage
\title{Level Two String Functions and Rogers Ramanujan Type Identities}

\author{Arel Genish and Doron Gepner}
\vskip20pt
\line{\it\hfill  Department of Particle Physics, Weizmann Institute, Rehovot, Israel\hfill} 

\abstract
The level two string functions are calculated exactly for all simply laced Lie algebras,
using a ladder coset construction.
These are the characters of cosets of the type $G/U(1)^r$, where $G$ is the algebra at level two and
$r$ is its rank. This coset is a theory of generalized parafermions.
A conjectured Rogers Ramanujan type identity is described for these characters.
Using the exact string functions, we verify the Rogers Ramanujan type expressions, that are the main focus of this work.
\endpage 

\mysec{Introduction}

One of the remarkable properties of two dimensional systems is the expression of many characters
of conformal field theories (CFT) as generalized Rogers Ramanujan type sums, which we denote as GRR. Many
such examples have been worked out in the literature
\REF\Lep{J. Lepowski and M. Primc, Contemporary Mathematics vol. 46 (AMS, Providence, 1985).}
\REF\Das{V.S. Dasmahapatra, T.R. Klassen, B.M. McCoy and E. Melzer, J. Mod. Phys. B7 (1993) 3617.}
\REF\Kedem{R. Kedem, T.R. Klassen, B.M. McCoy and
E. Melzer, Phys. Lett. 307B (1993) 68.}
\REF\Baver{E. Baver and D. Gepner, Phys. Lett. B372 (1996) 231.} 
\REF\Kun{A. Kuniba, T. Nakanishi and J. Suzuki, Mod. Phys. Lett. A8 (1993) 1649.}
\REF\BelGep{A. A. Belavin and D. R. Gepner, Lett. Math. Phys. 103 (2013) 1399.}
\r{\Lep,\Das,\Kedem,\Baver,\Kun,\BelGep}. Yet, the origin of these identities remains
somewhat mysterious.

Some of this mystery is explained by the deep connection to two dimensional solvable
lattice models, in particular, the so called, rigid solid on solid models (RSOS). This connection 
was first noted by Baxter, in the context of the Hard Hexagon model 
\REF\Baxter{R. J. Baxter, Exactly solved models in statistical mechanics, (Dover Book on Physics).}
\r\Baxter, and was considerably further developed in the works
\REF\Melzer{E. Melzer, Int. J. Mod. Phys. A 9 (1994) 1115.}
\REF\Lattice{D. Gepner, Phys. Lett. B 348 (1995) 377.}
refs. \r{\Melzer,\Lattice}.
The crux of this connection is that the local state probabilities in the RSOS models
are identical to the characters of the fixed point CFT in the appropriate regime.
(More precisely, the one dimensional configuration sum, which is the center piece of
the local state probability, is equal to the characters.) This connection leads to the conjecture that GRR identities exist
for every CFT that appears as a fixed point for some RSOS model in some regime, and
vice versa.

The connection with RSOS lattice models also implies a way to prove these identities \r{\Melzer,\Lattice}, where we consider
the finite one dimensional configuration sum, which obeys some simple recursion relations.
This sum is equal to a version of the characters, which is the finite GRR identities.

In this paper, we describe new GRR identities for the cosets of the type
$$H={G_2\over U(1)^r},\e$$
where $G$ is any simply laced algebra, at level two, and $r$ is its rank.
The case of $A_r$ was already described in ref. \r\BelGep, and we extend
this work to all the other algebras. The GRR expression for the identity characters for the models $H$
was conjectured in ref. \r\Kun, and we agree with this conjecture.
We find GRR identities for all the characters, where the weight in $G_2$ has
Dynkin index one.

We achieve these new GRR by calculating exactly the string function of $G_2$, which are the characters of $H$ (up to factors 
of Dedekind’s eta functions).
The calculation is done by utilizing a ladder coset construction, which is a product 
of intermediate cosets, where each of the latter can be solved exactly by identifying it
as a simpler CFT.

The connection with RSOS of our GRR identities, and, in particular, their proof in this way, 
is an interesting question, which we are currently investigating.

\mysec{The general conjecture}

Our objective is to describe GRR type equalities (Generalized Rogers
Ramanujan identities), for the characters of the cosets of the type 
$$H={G_2\over U(1)^r},\e$$
where $r$ is the rank of $G$ and $G$ is any of the simply laced 
algebras which are of the
type $A_r$, $D_r$, $E_6$, $E_7$ and $E_8$, at level two.
The case of $A_r$ was already discussed in ref. \r\BelGep,
where GRR identities were found for all the characters of $H$.
Our aim here is to generalize these results to all of the simply 
laced algebras.

The characters of the theory $H$, for any $G$ at any level, were discussed 
in ref. \REF\GepNew{D. Gepner,
Nucl. Phys. B 290 (1987) 10.}\r\GepNew,
and are related to generalized parafermions. Briefly, they are labeled by two weights,
$\Phi^\Lambda_\lambda$ where $\Lambda$ is a highest weight of $G$
at level $k$ (here $k=2$) and $\lambda$ is an element of the weight
lattice of $G$. The weights obey the condition $\Lambda-\lambda\in M$, where we denote
by $M$ the root lattice of $G$. To each of the fields $\Phi^\Lambda_\lambda$,
we associate a character $\chi^\Lambda_\lambda$, which is given
by the corresponding string function of $G_k$. The string functions are denoted by
$c^\Lambda_\lambda(\tau)$ and are defined as
$$c^\Lambda_\lambda(\tau)=\Tr_{{\cal H}^\Lambda_\lambda} e^{2 \pi i\tau (L_0-c/24)},\e$$
where $L_0$ is the dimension and ${\cal H}^\Lambda_\lambda$ is
the representation of the affine $G_k$ with highest weight $\Lambda$
and $U(1)^r$ charge $\lambda$.
The characters of the coset $H$ are then given by
$$\chi^\Lambda_\lambda(\tau)=\eta(\tau)^r c^\Lambda_\lambda(\tau),\e$$
where $\eta(\tau)$ denotes the Dedekind's eta function,
$\lambda$ is defined modulo $k M$ and $\Phi^\Lambda_\lambda=\Phi^\Lambda_{\lambda+\mu}$,
where $\mu\in k M$.

Let us present our main result which is a GRR expression for the 
characters of the theory $H$. We denote by $C_G$ the Cartan matrix of the algebra $G$ which can be any 
of the algebras $A_r$, $D_r$, $E_6$, $E_7$ and $E_8$. 
Let $\Lambda$ stand for any of the fundamental weights which
have Dynkin index $1$, $\Lambda=\Lambda_i$ or $\Lambda=0$, where
$i=1,2,\ldots,r$, and $\theta\Lambda_i=1$, where $\Lambda_i$ is one
of the fundamental weights and $\theta$ is the highest root.
We conjecture, that the characters of $H$ have the following expression,
$$\chi^{\Lambda}_\lambda(\tau)=q^{\Delta_\lambda^{\Lambda}}\sum_{\vec b=0\atop \vec b=\vec Q\mod2}^
\infty {q^{\vec b C_G \vec b^t/4-b_i/2}\over (q)_{b_1} (q)_{b_2}\ldots
(q)_{b_r}},\e$$
where $(q)_n$ is 
$$(q)_n=\prod_{m=1}^n (1-q^m),\e$$
and for $\Lambda=0$ we take $b_i=0$. $\Delta^{\Lambda}_\lambda$ is some dimension
which we do not specify. 
Here $q=e^{2\pi i \tau}$.
We use the vector notation
$\vec b=(b_1,b_2,\ldots, b_r)$. $Q$ is a vector of integers modulo 
$2$ defined by
$$\Lambda-\lambda=\sum_{k=1}^r Q_k \alpha_k\mod 2 M,\e$$
where $\alpha_k$ denotes the $k$th simple root\foot{Because
of the field identifications in the coset $H$, $\chi_{\sigma(\lambda)}^{\sigma(\Lambda)}(\tau)=
\chi^\Lambda_\lambda(\tau)$, where $\sigma$ is any automorphism of
the Dynkin diagram of the affine $\hat G$, at level two, it is enough to consider
only $\Lambda$ which are fundamental weights or zero.}. 
For the algebra $G=A_r$ this reproduces the result of ref. \r\BelGep.
For the case of $\Lambda=0$ this agrees with a conjecture which was
put forwards in ref. \r\Kun. 

The sum in eq. (2.4) consists of
generalized Rogers Ramanujan identities. To establish this GRR
we need to calculate the string functions of $G=D_r,E_6,E_7,E_8$
at level two. We could, in principle, do this numerically using
group theoretic recursion relations for the algebras in question.
This is, in fact, too difficult to do by a computer program.
So, instead, we will calculate the string functions exactly, 
finding closed analytical  
expressions for all the string functions.
We accomplish this by generalizing the ladder coset construction of ref. 
\r\BelGep. Once such expressions are found we can verify the GRR, eq. (2.4).

\mysec{The algebra $D_n$}

We start with $D_n=SO(2n)$ at level two. We can write the coset
$H=SO(2n)_2/U(1)^n$ as a product of cosets,
$$H=\prod_{r=1}^n {SO(2r)_2\over SO(2r-2)_2 \times U(1)},\e$$
where all the intermediate cosets cancel in the product.
Of course, care must be taken, to make this correct in conformal
field theory. It means that the same representation must be imposed when
the same group appears in the numerator and the denominator, and
this representation has to be summed over. In other words, for this product coset 
we take the special modular invariant,   
where all the intermediate representations are identical and 
are summed over. If we denote by $G^\Lambda_2$ the highest weight field   
$\Lambda$ for the group $G$ at level two, and by $G^\Lambda_2/H^\lambda_2$
the appropriate branching function for this coset, we can write the characters 
as,
$$\chi^\Lambda_\lambda(\tau)=\sum_{\mu_r,\ r=1,2,\ldots,n-1}
\prod_{r=1}^n {SO(2r)_2^{\mu_r}\over SO(2r-2)_2^{\mu_{r-1}} \times U(1)^{\lambda\alpha_r}},
\e$$
where $\mu_n=\Lambda$ and $\alpha_r$ is the $r$th simple root of $SO(2n)$. 

So, we proceed to calculate the characters of cosets of the type 
$$H_r={SO(2r)_2\over SO(2r-2)_2\times U(1)}.\e$$ 
It is easy to see that for any $r$ this coset has the central charge $c=1$. Thus, it has to be one of the
$c=1$ models which were completely classified in ref. \REF\Ginsparg{
P. H. Ginsparg, Nucl. Phys. B  295  (1988) 153.}
\r\Ginsparg.
By calculating the dimensions of the fields
in this coset we deduce that the coset $H_r$ is given by a $Z_2$ orbifold of one free boson at the 
radius $\sqrt{2 r(r-1)}$. This implies that all the dimensions in this coset are of the form
$$\Delta_m={m^2\over 4 r (r-1)},\e$$
where $m$ is any integer, or $\Delta=1/16,9/16$, which are the dimensions of the twist fields. Eq. (3.4)
can be easily checked directly, by calculating the dimensions of the fields in the coset $H_r$.

Now, using eq. (3.2), we wish to calculate the characters of the coset $H=SO(2n)_2/U(1)^n$.
The calculation is simplified since $H$ is equivalent to the theory which is 
a product of $n-1$ single boson orbifolds at various radii,
implying that $H$ is a $Z_2$ orbifold of $n-1$ free bosons propagating at some lattice $M$.
If we denote the bosons as $\phi_i$, $i=1,2,\ldots,n-1$, the orbifold is given by $\phi_i\rarrow-\phi_i$.
So it remains to find the lattice $M$. This lattice can be constructed inductively using the ladder
coset expression eq. (3.2). We use the beta method, ref.  
\REF\Space{
  D. Gepner,
Nucl. Phys. B 296, 757 (1988) 757.}\r\Space,
to account for the summation of the intermediate 
weights, to find that the lattice $M$ is the root lattice of the group $SU(n)$ scaled by
$\sqrt{2}$, i.e., $M\approx \sqrt{2} M_{SU(n)}$. This is termed \REF\KacBook{
Victor G. Kac, “Infinite-Dimensional Lie Algebras” [Paperback] (1994).}
the $SU(n)$ lattice at level two, and the characters are the classical theta functions of $SU(n)$ level two, ref. \r\KacBook.
Thus, we established:

{\rm \bf Theorem (1):}
The coset $SO(2n)_2/U(1)^{n}$ is equivalent to the theory of $n-1$ free bosons, $\phi_i$, $i=1,2,\ldots,n-1$, 
moving on the lattice
$\sqrt{2} M_{SU(n)}$, $Z_2$ orbifolded by the twist $\phi_i\rarrow -\phi_i$. 

\par
Let us demonstrate how the $SU(n)_2$ lattice is built. We do this 
inductively. First, for $SO(4)$, we consider the theory
$${SO(4)_2\over U(1)^2},\e$$
which has the central charge $1$ and is equivalent to a $Z_2$ 
orbifold of one free boson at the radius $2$. This is the $SU(2)$
torus scaled by $\sqrt2$, in accordance with the theorem.

To identify the lattice for $SO(2n)_2/U(1)^n$ we consider the 
character of the identity,
$$\chi^0_0(\tau)={SO(2n)_2^0\over (U(1)^n)^0},\e$$
where we denoted by $0$ the unit representation. Now, this character
is equal to a theta function on some lattice $M$, which we wish to
identify. So, the character is equal to
$$\chi^0_0(\tau)=\sum_{p\in M} e^{\pi i\tau p^2},\e$$
up to factors of Dedekind's eta function.

Let us show how this lattice is obtained for $SO(6)$.
We know that the character $\chi_0^0(\tau)$ may be written as
$$\chi_0^0(\tau)=\sum_\mu {SO(6)_2^0\over SO(4)_2^\mu\times U(1)^0}
\times {SO(4)_2^\mu \over SO(2)^0\times U(1)^0},\e$$
where we denoted the representation by the superscript.
For $\mu=0$, this is a product of two bosonic $c=1$ theories,
moving on the two dimensional lattice $N$, generated by the vectors $\{0,2\}$ and
$\{2\sqrt3,0\}$, as discussed above. Now, the lattice $N$ will be
enhanced, i.e., it will be a sub lattice of $M$, due to the
representations with $\mu\neq0$. To find $M$ it is enough to consider
$\mu=(1,1)$, which is the second fundamental weight of 
$SO(4)$. Thus we have to identify the momenta of the two fields 
$$P_1=SO(6)_2^0/ SO(4)_2^{(1,1)}\times U(1)^0,\e$$
and
$$P_2=SO(4)_2^{(1,1)}/SO(2)^0\times U(1)^0.\e$$
This we do by calculating their dimensions. The dimension of
$P_1$ comes out to be $3/2$, and that of $P_2$ is $1/2$.
Thus, they correspond to the beta vector $\beta=\{\sqrt3,-1\}$.
So, the lattice $M$ is the lattice generated by $\beta$ and $N$.
This is the lattice generated by $\{0,2\}$ and $\beta=\{\sqrt3,-1\}$.
Multiplying the scalar products of these two generators, we find
that the scalar products matrix is 
$$H=\pmatrix{ 4 &-2\cr -2 &4\cr},\e$$
which is twice the Cartan matrix of $SU(3)$. Thus $M$ is given by
$\sqrt2 M_{SU(3)}$, in agreement with theorem (1).

We can continue, in this fashion, for any $SO(2n)$ and we find that
$M$ is indeed $\sqrt2 M_{SU(n)}$, establishing theorem (1).  

Theorem (1) allows to calculate easily the string functions of $SO(2n)_2$, since they are given by
the characters of a simple $Z_2$ orbifold. 
The fields are labeled by $\lambda$, where $\lambda$ is a weight of $SU(n)$ and their dimensions are 
$$\Delta_\lambda={\lambda^2\over 4}.\e$$
In addition, there are the twist fields whose dimensions are $\Delta={n-1\over16},\ {n+7\over 16}$.
The characters are expressed by a level two $SU(n)$ classical theta
functions. These are defined by
$$\Theta_{\lambda,m}(\tau)=\sum_{\mu\in R+\lambda/m} q^{\pi i m \mu^2},\e$$
where $\lambda$ is any element of the weight lattice of $SU(n)$ and $R$ denotes the root lattice of $SU(n)$. 
We take the level to be $m=2$ which means that this is a level two theta function.

The characters of the $Z_2$ orbifold are given by standard argumentation and are divided into the twisted
and the untwisted sectors. In the untwisted sector we have three types of characters. The characters of the 
zero momenta fields are
$$q^{(n-1)/24} \chi_\pm(\tau)={1\over2}\Theta_{0,2}(\tau)/\prod_{k=1}^\infty (1-q^k)^{n-1}\pm
{1\over2}\prod_{k=1}^\infty (1+q^k)^{-n+1},\e$$
where the second term arises due to the sector twisted in the time direction.
The nonzero momenta do not have such a contribution and are given simply by
$$\chi_\lambda(\tau)={1\over n_\lambda}{\Theta_{\lambda,2}(\tau)\over \eta(\tau)^{n-1}},\e$$
where $\lambda$ is a weight of $SU(n)$, $n_\lambda=2$ when $\lambda$ is on the root lattice and 
$n_\lambda=1$, otherwise.
In the twisted sector, we have two fields which have the characters, $\chi_{t,\pm}(\tau)$, where
$$q^{(n-1)/24} q^{(1-n)/16}\chi_{t,\pm}(\tau)={1\over2}\prod_{k=1}^\infty (1-q^{k-1/2})^{-n+1}\pm
{1\over2}\prod_{k=1}^\infty (1+q^{k-1/2})^{-n+1}.\e$$
We can determine which weight of $SU(n)$ corresponds to each character by simply comparing the dimensions
(some care is required when there is more than one field of the same dimension).
Thus, we can confirm directly the GRR eq. (2.4). We will give some examples below.

We first consider the unit character in the coset $SO(2n)_2/U(1)^n$. It is
given by the $\lambda=0$ character in the corresponding $SU(n)_2$ orbifold. Thus we have
the following identity for any $n$,
$$\chi_{+}(\tau)=q^{(1-n)/24} \sum_{b_i=0\atop b_i=0\mod2}^\infty {q^{\vec b C_n \vec b^t/4}\over (q)_{b_1} (q)_{b_2}\ldots
(q)_{b_n}},\e$$
where $i$ goes from $1$ to $n$ and $\chi_+(\tau)$ was given by eq. (3.14). $C_n$ is the Cartan matrix of $D_n\approx SO(2n)$ and 
$\vec b=(b_1,b_2,\ldots, b_n)$. This is a beautifully simple identity of the Rogers Ramanujan type
that we
verified directly using a Mathematica program for $n=2$ to $n=8$ and up to order $q^{20}$.

Other particularly simple identities arise from the twisted sector. They correspond in the coset to the
characters $\chi^s_s(\tau)$ and $\chi^s_{s+\alpha_1}(\tau)$, in the notation of eqs. (2.2) and (2.3),
where $s$ stands for the spinor representation and 
$\alpha_1$ is the first simple root. The simple roots of $D_n$ are $\alpha_i=\epsilon_i-\epsilon_{i+1}$,
for $i=1,2,\ldots, n-1$ and $\alpha_n=\epsilon_{n-1}+\epsilon_n$, where $\epsilon_i$ are orthogonal unit vectors.
Then, we have the following GRR identities,
$$\chi_{t,+}(\tau)=q^{\Delta} \sum_{b_i=0\atop b_i=0\mod2}^\infty {q^{\vec b C_n \vec b^t/4-b_n/2}\over 
(q)_{b_1} (q)_{b_2}\ldots (q)_{b_n}},\e$$ 
where $\chi_{t,+}(\tau)$ was given by eq. (3.16). For the other twist field we have,
$$\chi_{t,-}(\tau)=q^{\Delta_1} \sum_{b_i=0\atop b_i=0\mod2\ b_1=1\mod2}^\infty {q^{\vec b C_r \vec b^t/4-b_n/2}\over 
(q)_{b_1} (q)_{b_2}\ldots (q)_{b_n}},\e$$
where $\Delta$ and $\Delta_1$ are some dimensions. The above identities agree with our general GRR conjecture, eq. (2.4).

Let us give another example. Consider the character $\chi^v_v(\tau)$, for the algebra $SO(8)$, 
where $v$ is the vector representation. 
The field has the dimension $3/16$. The dimension determines that it corresponds to the field in the $SU(4)_2$
bosonic theory, which has the weight $\lambda=\nu_1-\nu_2$, where $\nu_i$ is the $i$th fundamental weight of $SU(4)$.
Thus, the character is given by
$$\chi^v_v(\tau)=\Theta_{\nu_1-\nu_2,2}(\tau)/\eta(\tau)^3,\e$$
where we denoted with $\Theta_{\lambda,2}(\tau)$ the theta function on the $SU(4)$ lattice, at level two, defined
by eq. (3.13). According to our general GRR conjecture, eq. (2.4), we have the following identity,
$$\chi_v^v(\tau)=q^{\Delta} \sum_{b_i=0\atop b_i=0\mod2}^\infty {q^{\vec b C_4 \vec b^t/4-b_1/2}\over 
(q)_{b_1} (q)_{b_2}\ldots (q)_{b_4}},\e$$
where $C_4$ is the Cartan matrix of $D_4=SO(8)$, $\Delta$ is some dimension, 
and $b_1$ corresponds to the vector representation.
We verified directly this identity up to order $q^{20}$, using a Mathematica program, and indeed it is correct.
 
Other examples of this kind can be demonstrated. We checked all the identities for $\chi^\Lambda_\lambda(\tau)$,
where $\Lambda\theta=1$ and
any $\lambda$ using a Mathematica program for $r=2$ to $r=8$ up to order $q^{20}$, and indeed they all hold.
However, the condition $\Lambda\theta=1$, where $\theta$ is the highest root (which means that $\Lambda$ has a Dynkin index one), 
is not obeyed for some of the weights $\Lambda$ and, in this case, our conjecture
breaks down. Further work is needed to find the GRR for the weights $\Lambda$
such that $\Lambda\theta\neq1$.

\mysec{The algebra $E_6$}

Let us consider now the algebra $E_6$. The relevant coset here is, 
$$H=(E_6)_2/U(1)^6.\e$$
We can present this coset as 
$$H=P\times {SO(10)_2\over U(1)^5},\e$$
where the coset $P$ is
$$P^\Lambda_{\mu,q}={((E_6)_2)^\Lambda\over (SO(10)_2)^\mu\times U(1)^q},\e$$
where $SO(10)\times U(1)$ is a maximal subgroup of $E_6$. The characters are given by
$$\chi^\Lambda_{\lambda,q}(\tau)=\sum_\mu P^\Lambda_{\mu,q} \times {SO(10)^\mu \over (U(1)^5)^\lambda},\e$$
where we denoted, again, by $\chi^\Lambda_{\lambda,q}(\tau)$ the characters with, $\Lambda$, $\mu$ and  $q$,
the weights of the corresponding algebras.

Since we know the characters of $SO(10)_2$, it is enough to determine the characters of $P$. 
We note that the central charge of $P$ is 
$$c=8/7.\e$$
It is the same central charge as that of the parafermion theory $W=SU(2)_5/U(1)$ \REF\ZF{
V. A. Fateev and A. B. Zamolodchikov,
Sov. Phys. JETP 62 (1985) 215, [Zh. Eksp. Teor. Fiz. 89 (1985) 380].}\r\ZF.
Calculating the dimensions
of the fields in the coset $P$, reveals that some of them are, indeed, as in the coset $W$.
In fact, the theory $P$ is equivalent to the coset $W$ orbifolded by the symmetry $A\rarrow A^\dagger$,
where $A$ is any field in the theory. The dimensions of the fields in this orbifold were calculated, along with the
characters, in ref. \REF\GepPar{
D. Gepner and H. Partouche,
``On The Real Part of a Conformal Field Theory,''
[arXiv:1310.8514 [hep-th]].}\r\GepPar,
and there is a complete agreement of all the dimensions,
leading us to theorem (2).

{\rm \bf Theorem (2):}
The characters of the theory $(E_6)_2/U(1)^6$ are given by the characters of the orbifolded fifth parafermion
times the characters of $SO(10)_2/U(1)^5$, known from theorem (1), where there is summation
on the intermediate weight, eq. (4.4).

The central charge of the $k$th parafermion, $k=1,2,\ldots$, is given by
$$c={3k\over k+2}-1,\e$$
and the dimensions are given by
$$\Delta^l_m={l(l+2)\over 4(k+2)}-{m^2\over 4 k},\e$$
where $l=0,1,\ldots,k$ and $m=0,1,\ldots, k$ and $l=m\mod2$.
The characters are given by the string functions of $SU(2)$, up to a factor of Dedekind’s eta function,
and can be expressed as Hecke 
indefinite modular forms \REF\KP{
V. Kac, D. Peterson,
Adv. Math., 53 (1984), 125.}
\r\KP. For the fifth parafermionic theory,
we use the expression for the characters appearing in ref.  \r\KP\ (page 219).
Define 
$$\epsilon(m,n)=\exp\left(2\pi i {m-n+2\over8}\right),\qquad n=0\mod 2,\e$$
$$\epsilon(m,n)=\exp\left( 2\pi i {n+1\over4}\right),\qquad n=1\mod2.\e$$
Then, the characters of the fifth parafermion are given by
$$d^p_r(\tau)=(-1)^p \eta(\tau)^{-2} \sum_{m,n=-\infty \atop {m=r\mod5\atop {n=2p+2\mod7\atop 7m^2+5n^2=4\mod16}}}
^\infty \epsilon(m,n) q^{(7m^2+5n^2)/560},\e$$
where $p=r\mod2$. 

Due to the orbifold, we have additional twist (disorder) fields. Their dimensions are, 
ref. \r\GepPar,
$$\Delta_\gamma={(\gamma+p/2)^2 \over 4(k+2)}-{1\over 48}+{c\over 24},\e$$
where $k$ is the level, here $k=5$, and $p=k\mod 2$, $p=0$ or $1$. The label $\gamma=0,1,\ldots k+1$,
and the central charge $c$ was given by eq. (4.6).

It can be verified that all the fields in the coset $P$ have dimensions which correspond either to the 
untwisted sector, eq. (4.7), or the twisted sector, eq. (4.11), in agreement with theorem (2).

The characters of $W$, orbifolded by the twist $A\rarrow A^\dagger$, were given in ref. \r\GepPar. In the untwisted sector we have two types
of characters.
$$y^l_m(\tau)=d^l_m(\tau),\e$$
for $l=0,1,\ldots,k$ and $m=1,2,\ldots,k-1$. I.e., the characters with $m\neq0$ are the same as in the
untwisted theory.
For $m=0$ we have
$$y^l_{0,\pm}(\tau)=d^l_0/2\pm B^l(\tau)/2,\e$$
where 
$$B^l(\tau)=\eta(2\tau)^{-1} \sum_{n=-\infty}^\infty (-1)^{nk} e^{2(k+2) \pi i \tau (n+{l+1\over 2(k+2)})^2}.\e$$
In the twisted sector, the characters are given by
$$d_{t,\pm}^\gamma(\tau)=w_{t}^\gamma(\tau)/2 \pm w_t^\gamma(\tau+1)/2,\e$$
where
$$w_{t}^\gamma(\tau)=\eta(\tau/2)^{-1} \left[\theta_{\gamma+p/2,k+2}(\tau)- \theta_{\gamma+p/2+k+2,k+2}(\tau)\right],\e$$
and where $\theta_{l,k}(\tau)$ stands for the $SU(2)$ classical theta function,
$$\theta_{l,k}(\tau)=\sum_{j=-\infty}^\infty e^{2\pi i k\tau (j+{l\over 2k})^2},\e$$
and $l$ is any integer.

Let us give some examples. We take the basis for the simple roots of $E_6$
which corresponds to the breaking of $E_6$ to $SO(10)\times U(1)$,
\def\h{-{1\over2}}
$$\pmatrix{\alpha_1 \cr  \alpha_2 \cr  \alpha_3 \cr  \alpha_4 \cr  \alpha_5 \cr \alpha_6\cr}=
\pmatrix {1 &  -1 & 0 & 0 & 0&  0 \cr 
          0 &  1 &  -1 &  0 & 0 & 0\cr
          0 &  0 &  1 & -1 & 0 &   0\cr
          0 & 0 & 0 & 1 & 1 & 0\cr
          \h & \h & \h & \h & \h & {\sqrt3\over2} \cr
          0 &  0 &  0 &   1 & -1 & 0\cr},\e$$
where we denoted by $\alpha_i$ the $i$th simple root.  
The simplest fields to consider in the $P$ coset are
$$(E_6)_2^0\over SO(10)_2^\lambda\times U(1)^{-\sqrt{3}/2},\e$$
where $\lambda=s$ or $v+ \bar s$, 
where $s$ ($\bar s$) denotes the spinor (anti spinor) representations and $v$ is the vector representation of $SO(10)$.
The upper index is the representation, which, here, is the singlet of $E_6$. These fields have the dimensions $1/4$ or $3/4$, and are thus fields in the
twisted sector with $\gamma=2$, $\Delta_\gamma=1/4,3/4$, eq. (4.11). There are two characters for these fields which are
$d_{t,\pm}^2$, eq. (4.15), corresponding respectively to $\lambda=s$ or $\lambda=v+\bar s$. These fields are multiplied 
by the fields in the $SO(10)_2/U(1)^5$ theory, which are the twist fields of the $SU(5)_2$ orbifold, whose characters are given by eq. (3.16),
$$u_{\pm}(\tau)=q^{-4/24} q^{4/16}\left({1\over2} \prod_{j=0}^\infty (1-q^{j+1/2})^{-4}\pm {1\over2}\prod_{j=0}^\infty (1+q^{j+1/2})^{-4}
\right).\e$$
Thus, the two characters of the coset
$${(E_6)_2^0\over (U(1)^6)^\zeta},\e$$
where  $\zeta$ is either the weight $\zeta_+=\{1/2,1/2,1/2,1/2,1/2,-\sqrt3/2\}$ or the weight
$\zeta_-=\{3/2,1/2,1/2,1/2,-1/2,-\sqrt3/2\}$ are given by
$$r_{+}(\tau)=u_+(\tau) d_{t,+}^2(\tau)+u_-(\tau)d_{t,-}^2,\e$$
and
$$r_-(\tau)=u_+(\tau) d_{t,-}^2(\tau)+u_-(\tau) d_{t,+}^2(\tau).\e$$
We observe, now,  that these two characters are given by the GRR type identity
$$r_{\pm}(\tau)=\sum_{\vec b=0\atop \vec b=0\mod 2\ b_5=e\mod2}^\infty {q^{\vec b C_{E6} \vec b^t/4}\over
(q)_{b_1} (q)_{b_2}\ldots (q)_{b_6}},\e$$
where $e=0$ for $r_+$, $e=1$ for $r_-$ and $C_{E6}$ stands for the Cartan matrix of $E_6$.
We checked these two identities, by a Mathematica program, up to order $q^{20}$, and indeed they hold.  
 
Let us give another example. Consider the character,
$$b={(E_6)_2^{\mu_1}\over (U(1)^5)^{\lambda_3}\times U(1)^{1/\sqrt3}},\e$$
where the superescript denotes the representation and
where $\mu_1$ is the first fundamental weight of $E_6$ (the $27$) and $\lambda_3$ is the third fundamental
weight of $SO(10)=D_5$. Using theorem (2), this string function can be written as
$$b=\sum_\lambda {E_6^{\mu_1}\over SO(10)^\lambda \times U(1)^{1/\sqrt3}} \times {SO(10)^\lambda\over (U(1)^5)^{\lambda_3}  },\e$$
where the sum is over $\lambda$ which are equal to $\lambda=\lambda_1,\lambda_3,\lambda_5$, where $\lambda_i$
denotes the $i$th fundamental weight of $SO(10)$.
We identify each of the sub characters in eq. (4.26) according to the dimensions, which are
$$b= [{3\over 35}] \times [7/10]+[{17\over 35}]\times[ 3/10]+[{2\over 7}]\times [1/2],\e$$
where $[x]$ denotes the character of the corresponding field with the dimension $x$.
We used the following expressions for the dimensions of the fields.
The dimension of the field $(E_6)_2^\mu/SO(10)_2^\lambda\times U(1)^q$ is
$$\Delta_{\lambda,q}^\mu=\mu(\mu+2\rho)/28-\lambda(\lambda+2\rho_{10})/20-q^2/4,\e$$
where $\rho$ is the sum of $E_6$ fundamental weights and $\rho_{10}$ that of $SO(10)$.
The dimension of $SO(10)_2^\lambda/(U(1)^5)^\omega$ is
$$\Delta^\lambda_\omega=\lambda(\lambda+2\rho_{10})/20-\omega^2/4.\e$$
This translates to the following character formula,
$$b=d^2_2 \Theta(\nu_1+\nu_3)  + d^3_1\Theta(\nu_1-\nu_3)+ d^2_0\Theta(\nu_1+\nu_4),\e$$
where we denoted by $\nu_i$ the $i$th fundamental weight of $SU(5)$ and
where $\Theta(\nu)$ is the classical theta function of $SU(5)$ at level two and the weight $\nu$, eq. (3.13),
$$\Theta(\nu)=\Theta_{\nu,2}(\tau)/\eta(\tau)^4,\e$$
which expresses the string functions of $SO(10)_2$, in accordance with theorem (1).
Finally, we can verify the following GRR in accordance with the conjecture, eq. (2.4),
$$b=q^\Delta \sum_{\vec b\atop {\vec b=0\mod2\atop  b_2=b_3=1\mod2}}^\infty {q^{\vec b C_{E_6} {\vec b}^t/4-b_1/2}\over
(q)_{b_1}\ldots (q)_{b_6}},\e$$
where $\Delta$ is some dimension.
We verified, by Mathematica, this GRR up to order $q^{20}$ and indeed it holds.

\mysec{The algebra $E_7$}

Let us turn now to the string functions of $E_7$ at level two. The relevant coset here is
$$P={(E_7)_2\over U(1)^7}.\e$$
Here we use the following ladder coset construction,
$$P=G\times {SO(12)_2\over U(1)^6},\e$$
where $G$ is the coset
$$G={(E_7)_2\over SO(12)_2\times U(1)},\e$$
and where there is a summation over the intermediate representations,
$$\chi^\Lambda_\lambda(\tau)=\sum_\mu {(E_7)_2^\Lambda\over SO(12)_2^\mu\times U(1)^{\lambda^\prime}} 
\times {SO(12)_2^\mu \over (U(1)^6)^{\bar \lambda}},\e$$
where $\lambda^\prime$ and $\bar\lambda$ are the projections of $\lambda$ to the $U(1)$ and $U(1)^6$,
respectively.

We can calculate the central charge of $G$, and it is given by $c=13/10$. We note that it is the same central
charge as the product of minimal models $M_3\times M_5$, where $M_3$ is the Ising and $M_5$ is the three state 
Pott's model,
whose central charge is $c=1/2+4/5=13/10$. Calculating the dimensions in the coset, we note that
the dimensions are given by the sum of the dimensions of the two minimal models. Thus, we establish that
$G$ is given by $M_3\times M_5$. Since we know the string functions of $SO(12)_2$ we have a closed formula
for the string functions of $(E_7)_2$, which are the characters of the coset $(E_7)_2/U(1)^7$.
This is expressed in the following theorem (3).

{\rm \bf Theorem (3):}
The coset $(E_7)_2/U(1)^7$ is equivalent to the conformal field theory which is the product of the two minimal
models $M_3\times M_5$, times the orbifold of free bosons on $SU(12)_2$, where the product is according
to eq. (5.4).

The unitary minimal model $M_k$, refs.
\REF\BPZ{
A.A. Belavin, A.M. Polyakov, A.B. Zamolodchikov,
Nucl. Phys., B241 (1984), 333.} 
\REF\RC{A. Rocha-Caridi, In: J. Lepowsi, S. Mandelstam, I. M. Singer (eds.),
Vertex operators in mathematics and physics, Springer, New York (1984).}
\r{\BPZ,\RC},
has the central charge
$$c=1-{6\over k(k+1)},\qquad k=3,4,\ldots,\e$$
and the dimensions of the primary fields are,
$$\Delta_{n,m}={(n (k+1)-m k)^2-1\over 4 k(k+1)},\e$$
where $n=1,2,\ldots, k-1$ and $m=1,2,\ldots, k$. 
The characters of the primary fields can be expressed in terms of $SU(2)$ classical theta functions,
$$\chi_{n,m}^{(k)} (\tau)={\theta_{n(k+1)-m k,k(k+1)}(\tau)-\theta_{n(k+1)+m k,k(k+1)}(\tau)\over \eta(\tau)},\e$$
where the classical SU(2) theta functions, $\theta_{r,s}(\tau)$ were defined by eq. (4.17).
The dimension of the field in the $G$ theory
$$\Phi^\Lambda_{\mu,q}={(E_7)_2^\Lambda\over (SO(12)_2)^\mu\times U(1)^q},\e$$
where $\Lambda$ is a highest weight of $E_7$ at level $2$, $\mu$ is a weight of $SO(12)_2$, $q$ is the 
weight of the $U(1)$,
is given by
$$\Delta_{\mu,q}^\Lambda={\Lambda(\Lambda+2\rho)\over 40}-{\mu(\mu+2\rho_{12})\over 24}-
{q^2\over 4},\e$$
where $\rho$, or $\rho_{12}$, are the sum of the fundamental weights of $E_7$, or $SO(12)$, respectively.
We take a basis for the simple weights of $E_7$, which corresponds to the breaking into $SO(12)\times U(1)$,
\def\h{{1\over2}}
$$\pmatrix{\alpha_1 \cr \alpha_2\cr \alpha_3\cr \alpha_4\cr \alpha_5\cr\alpha_6\cr\alpha_7\cr}=
\pmatrix{
  1& -1& 0& 0& 0& 0& 0\cr 
  0& 1& -1& 0& 0& 0& 0\cr
  0& 0& 1& -1& 0& 0& 0\cr 
  0& 0& 0& 1& -1& 0& 0\cr
  0& 0& 0& 0& 1& 1& 0\cr 
  -\h&-\h&-\h&-\h&-\h&-\h&\sqrt2/2\cr 
  0& 0& 0& 0& 1& -1& 0\cr}.\e$$

Let us give an example. Consider the fields in the $(E_7)_2/U(1)^7$ theory which are
$$S_{\pm}={(E_7)_2^0\over (U(1)^7)^{\nu_\pm}},\e$$
where $\nu_+=-\alpha_6$ for $S_+$ and $\nu_-=-\alpha_6+\{1,0,0,0,0,-1\}$ for $S_-$, and
where $\alpha_6$ is the sixth fundamental root of $E_7$, $-\alpha_6=\{1,1,1,1,1,1,-\sqrt2\}/2$. 
According to theorem (3) the character of the fields $S_\pm$ can be written as
$$ S_{\pm}(\tau)=\sum_\mu {(E_7)_2^0\over SO(12)_2^\mu\times U(1)^{-\sqrt2/2}}\times
{SO(12)_2^\mu\over (U(1)^6)^{\kappa_\pm}},\e$$
where $\kappa_+=s$ and $\kappa_-=\bar s+v$ and $s$ is the weight of the spinor representation, $\bar s$ is the anti spinor
and $v$ is the vector, all of the algebra $SO(12)$. $\mu$ is summed over the two values $\mu_+=s$ and $\mu_-=v+\bar s$.
We can identify the fields in the $G$ theory, 
$$g_\pm={(E_7)_2^0\over SO(12)_2^{\mu_\pm} \times U(1)^{-\sqrt2/2}},\e$$
by determining  the dimensions of these fields, using eq. (5.9,5.6), which are $3/16$ and $27/16$, 
we infer that they correspond to the fields in the
minimal models $M_3\times M_5$, which are
$(1,2,1,2)$ and $(1,2,1,4)$, for $g_+$ and $g_-$, respectively, where we denoted by $(n_1,m_1,n_2,m_2)$ the
field in the product of minimal models $M_3\times M_5$, with dimensions $\Delta_{n_1,m_1}$ and $\Delta_{n_2,m_2}$, 
respectively. The two characters of these fields are given by eq. (5.7) and are
$$t_\pm(\tau)=\chi_{n_1,m_1}^{(3)}\chi_{n_2,m_2}^{(5)},\e$$
where $(n_1,m_1)$ and $(n_2,m_2)$ take the two values indicated before.
In accordance with theorem (3), these characters are multiplied by fields in the bosonic orbifold
$SU(6)_2$. Since the representations of $SO(12)_2$ are spinorial, in this example, these fields are the
two twist fields of the orbifolded $SU(6)_2$ theory, which are given by
$$q^{-5/16+5/24} u_\pm(\tau)= {1\over2}\prod_{m=1}^\infty [1-q^{m-1/2}]^{-5}\pm {1\over2}\prod_{m=1}^\infty
[1+q^{m-1/2}]^{-5}.\e$$
Thus, the two characters $S_\pm(\tau)$ are given by
$$\eqalign{r_+&=t_+ u_++t_- u_-,\cr
           r_-&=t_- u_++t_+ u_-.\cr}\e$$
$r_\pm$ are the corresponding exact characters of the coset $(E_7)_2/U(1)^7$. Thus we can check our GRR conjecture
eq. (2.4), for this specific case. We find the following identity, 
$$ q^{y_\pm} r_{\pm}(\tau)=\sum_{\vec b=0\atop \vec b=0\mod 2,\ b_1=1\mod2,\ b_6=e\mod2}^\infty 
{q^{\vec b C_{E7} \vec b^t/4}\over
(q)_{b_1} (q)_{b_2}\ldots (q)_{b_7}},\e$$
where $e=0$ for $r_+$ and $e=1$ for $r_-$, and $C_{E7}$ is the Cartan matrix of $E_7$. $y_\pm$ are some dimensions.
We checked
these two identities up to order $q^{20}$, by Mathematica, and indeed they hold. 

\mysec{The algebra $E_8$}

The remaining algebra is $E_8$. Our objective is to calculate the string functions of $E_8$ at level
two. To do this, we use the following ladder coset construction,
$$H^\Lambda_\lambda={(E_8)_2^\Lambda\over (U(1)^8)^\lambda}=\sum_\mu {(E_8)_2^\Lambda\over SO(16)_2^\mu}\times
{SO(16)_2^\mu\over (U(1)^8)^\lambda},\e$$
where, as usual, we denoted by the superscripts the highest weights of the representations.
Since, we know from theorem (1) the string functions of $SO(16)_2$ it is enough
to resolve the coset
$$G={(E_8)_2\over SO(16)_2}.\e$$
The central charge of this coset is calculated to be $c=1/2$. Since the theory $G$ is
unitary, it has to be equivalent to the Ising model, previously denoted as $M_3$. 
This yields an analytic expression for the string functions of $(E_8)_2$, and theorem (4) follows.

{\rm \bf Theorem (4):}
The conformal field theory, the coset $E_8/U(1)^8$ at level two, is equivalent to a product of the Ising model times
the orbifold of free bosons on the $SU(8)$ torus at level two.

In view of theorem (4), we can calculate the string functions of $(E_8)_2$ and verify our GRR conjecture,
eq. (2.4). Let us give an example. Consider the field in the $H$ theory,
$H^0_{\mu}$ where $\mu=s$ or $\mu=\bar s+v$, and where we denoted by $s$, $\bar s$ and $v$ 
the spinor, anti spinor and vector representations of $SO(16)$. In the ladder coset construction, eq. (6.1), this translates
to the product of characters
$$H^0_\mu(\tau)=G\times R_\pm,\e$$
where
$$G={(E_8)_2^0\over SO(16)^s},\e$$
and
$$R_\pm={SO(16)_2^s\over (U(1)^8)^\mu},\e$$
where we took $R_+$ for $\mu=s$ and $R_-$ for $\mu=v+\bar s$.
Since $G$ is a field in the Ising model, it can be identified by its dimension, which is
$1/16$, where we used eq. (5.6). Thus, $G=\chi^{(3)}_{2,2}$, which is the character of the spin field in the Ising model.
Now, $R_\pm$ are given by the two characters of the twist fields of the bosonic $SU(8)_2$, which are
$$q^{-7/16+7/24} R_\pm(\tau)={1\over2} \prod_{j=1}^\infty (1-q^{j-1/2})^{-7}\pm
{1\over2} \prod_{j=1}^\infty (1+q^{j-1/2})^{-7}.\e$$
Thus, eq. (6.6) yields these two characters, enabling the verification of our GRR conjecture, eq. (2.4). 
We have the expression,
$$H_{\mu}(\tau)=q^{z_\pm} \sum_{\vec b=0\atop \vec b=Q_\pm \mod2}^\infty 
{q^{\vec b C_{E8} \vec b^t/4}\over
(q)_{b_1} (q)_{b_2}\ldots (q)_{b_8}},\e$$
where 
$Q_+=(1,0,0,0,0,0,0,0)$ for $R_+$ and $Q_-=(1,0,0,0,0,1,0,0)$ for $R_-$,
where $Q_\pm$ is defined modulo two
and
where $C_{E_8}$ is the Cartan matrix of $E_8$ and $z_\pm$ are some dimensions. 
Eq. (6.7) agrees with our general GRR conjecture, eq. (2.4). We verified directly
these two identities and, indeed, they hold up to order $q^{10}$.
  
The reader might wonder about the result we would get if we take in the GRR, eq. (2.4), the summation on $Q\mod2$, other than the
ones given by eq. (2.6). The result is surprising. For any choice of $Q\mod2$, we find that the GRR gives
one of the characters of the corresponding coset. This amounts to a great deal of interesting
and nontrivial identities, and it generalizes the identity by Slater, ref. 
\REF\Slater{L. J. Slater, Proc. Lond. Math. Soc. 54 (1953) 147.}\r\Slater,
for the Ising model. These identities we term 
generalized Slater identities. Some of these identities were already noted in ref. \r\BelGep.   

\mysec{Discussion}

In this paper, we described generalized Rogers Ramanujan (GRR) identities for the string functions of all the simply laced algebras at level two.
These are the characters of the conformal field theory (CFT) of generalized parafermions \r\GepNew, specialized to the appropriate algebra and level.
We find identities for all the characters, such that the weight in the algebra has Dynkin index one. 
These identities are surprisingly simple and pretty, and they center around the Cartan matrix of the algebra in question, in a type
of ADE classification for such identities. This adds considerably to the GRR identities which were already established
for characters of certain conformal field theories
\r{\Lep,\Das,\Kedem,\Baver,\Kun,\BelGep}.

An interesting
question is how to extend these identities to the weights which have Dynkin index greater than one.
Another important question is how to extend these GRR identities to the levels which are greater than two and to the non simply laced 
algebras.

Perhaps, the most interesting issue, from the physics point of view, is the relation to RSOS lattice models.
For the case of $A_r$ these models were investigated by Jimbo et al. 
\REF\Jimbo{M. Jimbo, T. Miwa, M. Okado,
Nucl. Phys. B 300 [FS22] (1988) 74.}\r\Jimbo.
The relation between GRR and RSOS lattice models was established before in a great deal of  cases \r{\Baxter,\Melzer,\Lattice,\Baver}.
Importantly, this relation, once established, will furnish a way to prove the GRR identities,
described here, and will also provide an expression for the finite one dimensional configuration sum in the RSOS models,
which is the main piece of the local state probabilities (or, the one point functions),
a central problem in this field.

\ack

We thank Ida Deichaite for remarks on the manuscript.

\refout
\bye